\documentstyle[12pt,epsf]{article}

\begin{document}

\def\gsim{\:\raisebox{-0.75ex}{$\stackrel{\textstyle>}{\sim}$}\:}
\def\lsim{\:\raisebox{-0.75ex}{$\stackrel{\textstyle<}{\sim}$}\:}
\renewcommand{\thefootnote}{\fnsymbol{footnote}}

\pagestyle{empty}
\begin{flushright} 
TUM-HEP-505/03 \\
April 2003
\end{flushright}

\vspace*{6mm}

\begin{center}

{\Large\bf Particle Physics Explanations for Ultra High Energy Cosmic Ray
Events}\footnote{Invited plenary talk at PASCOS03, Mumbai, India,
January 2003} \\
\vspace*{4mm}
{\large Manuel Drees} \\
\vspace*{2.5mm}
{\it Physik Department, TU M\"unchen, James--Franck--Str., 85748
Garching, Germany} 

\end{center}

\vspace*{1cm}
\begin{abstract}
The origin of cosmic ray events with $E \gsim 10^{11}$ GeV
remains mysterious. In this talk I briefly summarize several proposed
particle physics explanations: a breakdown of Lorentz invariance, the
``$Z-$burst'' scenario, new hadrons with masses of several GeV as
primaries, and magnetic monopoles with mass below $10^{10}$ GeV as
primaries. I then describe in a little more detail the idea that these
events are due to the decays of very massive, long--lived exotic
particles. 
\end{abstract}

\clearpage
\pagestyle{plain}
\section{Introduction}

The observation \cite{md1} of cosmic ray (CR) events with energy $E
\gsim 10^{11}$ GeV (ultra--high energy, UHE) poses at least two
distinct problems: \begin{itemize}

\item {\bf The energy problem:} It is very difficult, if not
impossible, to accelerate protons to such energies in any known
astrophysical source \cite{md2}.

\item {\bf The propagation problem:} Protons with $E \geq 5 \cdot
10^{10}$ GeV can photoproduce pions on the cosmic microwave background
(CMB). The GZK effect \cite{md3} implies that protons with $E \geq
10^{11}$ GeV should have been produced within a few dozen Mpc of
Earth. The same conclusion holds for photons [which get absorbed via
$\gamma_{\rm UHE} + \gamma_{\rm radio} \rightarrow e^+ e^-$, where
$\gamma_{\rm radio}$ belongs to the (extra)galactic radio background]
and heavier nuclei (which are broken up by collisions with CMB
photons). Current estimates of (extra)galactic magnetic field
strengths imply that protons with $E > 10^{10}$ GeV should point back
to their source if they are produced within a few dozen Mpc. There
are, however, no known near sources of high--energy particles in the
direction of the most energetic events.\footnote{There have been
claims \cite{md4} of statistically significant correlations between
the arrival directions of UHECR events and certain kinds of active
galactic nuclei (AGN), thought to be possible sources of UHE
particles. However, an independent analysis did not confirm this
\cite{md5}.}

\end{itemize}
Many suggested solutions of ``the UHECR puzzle'' actually only address
the second problem. I will briefly review these ideas in the following
section. In sec.~3 I will discuss two approaches that can solve both
problems, the main emphasis being on decays of super--massive
particles, before concluding in sec.~4.

\section{Partial solutions}

\subsection{Breaking Lorentz invariance}

Calculations of the GZK effect assume that $\sigma(\gamma p
\rightarrow N \pi) \ (N = n,p)$ is the same for $E_{\gamma} \simeq
E_{\rm CMB} \sim 10^{-3}$ eV, $E_p \sim 10^{11}$ GeV and $E_{\gamma}
\sim 200$ MeV, $E_p = m_p$. This assumption may be wrong if Lorentz
symmetry is violated. This is actually quite an old idea \cite{md6},
which has been rediscovered recently \cite{md7}. By assuming different
limiting velocities for different species of particles one can arrange
for a suppression of the GZK effect; one can even avoid it
altogether. In this case one would expect a completely smooth spectrum
around the GZK cut--off. In particular, one would not expect a bump
just below the GZK energy, which seems to be present in both the AGASA
and HiRes results \cite{md1}. Besides, it seems to me a step backwards
to give up one of the basic symmetries underlying our understanding of
Nature; it rather reminds me of Bohr's willingness to give up energy
conservation to explain nuclear $\beta$ decay spectra. Back then,
Pauli's neutrinos saved the day; I'm confident that the puzzle of the
UHECR events can also be solved without giving up well--established
principles. 

\subsection{Exotic hadrons as UHECR primaries}

If an exotic hadron $h_E$ with mass $m_{h_E} = r m_p$ is \cite{md7a}
the primary of the UHECR events, the GZK ``cut--off'' (better:
spectral break) will be pushed upwards by a factor $r$. Even the AGASA
data can be explained if $r \geq 4$. On the other hand, if $h_E$ is
too heavy, it looses energy too slowly in the air shower it initiates
when hitting the Earth's atmosphere. It has been estimated \cite{md8}
that the produced shower will only look sufficiently proton--like if
$r \lsim 50$.

There are at least two problems with this scenario. First,
acceleration becomes much more difficult to explain. Producing UHE
$h_E$ particles by first accelerating protons and then colliding them
with ambient matter (or photons) will likely over--produce neutrinos
and UHE photons, since the (inclusive) cross section for pion
production is several orders of magnitude larger than the cross
section for $h_E$ production; of course, this would also assume that
protons can be accelerated to energies well beyond the most energetic
observed CR event. Directly accelerating stable $h_E$ particles is
also problematic, since at least on Earth they only constitute at most
a tiny fraction of all matter, so the heavenly accelerator would
presumably also ``waste'' most of its energy on accelerating ordinary
protons and electrons. The second problem is that $h_E$ particles,
being strongly interacting and rather light, should have been produced
abundantly in collider experiments; it is difficult to believe that
they could have escaped detection.

\subsection{$Z$ bursts}
\setcounter{footnote}{0}

The idea \cite{md9} is that UHE neutrinos are produced somewhere;
since they can propagate almost freely through the Universe, this
neutrino source may be anywhere within one Hubble radius, 3 Gpc, or
so. Within the GZK radius around the Earth, a small fraction of these
UHE neutrinos annihilate on relic background antineutrinos to produce
$Z$ bosons, most of which decay hadronically, giving rise to the
observed UHECR events; of course, UHE $\bar \nu$ annihilating on relic
$\nu$ also contribute.

One problem with this explanation is that it aggravates the energy
problem by several orders of magnitude. One will need UHE neutrinos of
at least five times the energy of the most energetic UHECR event,
since $Z$ bosons decay into typically $\gsim 20$ hadrons. In order to
produce such neutrinos, one needs protons whose energy is higher by at
least another factor of five or so, i.e. one would need a source of
protons extending to $E \gsim 10^{13}$ GeV. Moreover, the flux
$\Phi_{\nu,{\rm UHE}} \gg \Phi_{\rm exp. \, UHECR}$, i.e. the intensity
of this source must be {\em much} higher than that needed to directly
accelerate the observed UHECR flux.\footnote{Such a source could
easily over--produce protons just below the GZK cut--off.} Indeed,
the required UHE neutrino flux is at best marginally compatible with
existing limits \cite{md10}.

\section{Complete solutions}

\subsection{Magnetic monopoles as UHECR primaries}

This is another old idea \cite{md11} that has been rediscovered
recently \cite{md11a}. The main observation is that magnetic monopoles
can actually be accelerated (rather than only deflected) by
(extra)galactic magnetic fields. Indeed, this acceleration is quite
efficient, partly due to the large effective charge ($\geq
1/\alpha_{\rm em}$) of the monopoles. According to current estimates
energies beyond $10^{11}$ GeV are easily accommodated in this way. Of
course, monopoles are stable, so they can have been produced in the
very early Universe, most plausibly during a phase transition. The
density, or flux, of monopoles required to explain the observed UHECR
events is safely below all known bounds \cite{md12}, as long as these
monopoles do not catalyze nucleon decay.

The biggest difficulty of this model is to explain why the observed
events look so much like proton--induced air showers. In order to
produce a shower at all, the monopole must be ultra--relativistic,
i.e. $m_M < 10^{10}$ GeV. These monopoles can therefore not be
associated with a Grand Unified symmetry; instead one has to postulate
the existence of an ``intermediate'' scale ${\cal O}(m_M)$. Of course,
direct monopole searches at colliders imply $m_M \gg m_p$. Simple
kinematics then implies that a monopole--induced air shower will
penetrate much more deeply into the atmosphere than a $p-$initiated
shower does, unless the cross section for inelastic scattering of a
monopole on an air nucleus is much {\em larger} than the corresponding
cross section for protons. The authors of ref.\cite{md12} argue that
bound states of monopoles carrying nontrivial $SU(3)_c$ charge might
conceivably have this property. In this view, successive collisions
could increase the length of the color--magnetic QCD string (or flux
tube) between the constituents, and hence the area of the bound
state. Note that this string can only break into
monopole--antimonopole pairs. Since monopoles are heavy, a purely
color--magnetic string might therefore store many orders of magnitude
more energy, and could correspondingly be many orders of magnitude
longer, than the more familiar color--electric string does, which can
break up into light $q \bar q$ pairs. The effective monopole--air
cross section can grow quickly through this mechanism, leading to the
desired short air shower. Note, however, that the monopoles, being
stable, should still reach the Earth (with non--relativistic
velocity); detecting these monopoles would be the ultimate test of
this explanation. 

\subsection{Decaying superheavy particles}
\setcounter{footnote}{0}

Another exotic possibility is that the UHECR events are due to the
decay (or annihilation) of some super--heavy long--lived $X$ particle
\cite{md13}. We obviously need $m_X \gsim 10^{12}$ GeV in order to
generate a significant flux of UHECR primaries at $E \geq 3 \cdot
10^{11}$ GeV. Just as obviously, the lifetime of $X$ must be at least
comparable to the age of the Universe, $\tau_X \gsim 10^{10}$
yrs. This immediately raises two particle physics questions: how can
such massive particles be so long--lived, with $\Gamma_X / m_X \lsim
10^{-54}$? And how could such massive particles ever have been
created?

One way to ensure the longevity of the $X$ particles is to embed them
into topological defects. The original proposal \cite{md13} envisioned
annihilating monopole--antimonopole pairs as source of UHECR
events. This idea has later been refined to the concept of a ``cosmic
necklace'', (anti)monopoles strung along a cosmic string
\cite{md14}. Cosmic strings themselves become unstable when they
intersect. Such kinds of defects could be formed during a phase
transition via the Kibble mechanism \cite{md15}.

Note that we now need the scale of symmetry breaking to be $\gsim
10^{12}$ GeV (as opposed to $\lsim 10^{10}$ GeV in the previous
subsection). This may be problematic, since with the scale of
(four--dimensional) inflation determined to be $\sim 10^{13}$ GeV, it
is quite possible that the post--inflationary Universe never was hot
enough to have been in the ``unbroken'' phase, which would require $T
\gsim 10^{12}$ GeV. Moreover, the increasingly well measured CMB
anisotropies do not seem to show any evidence that topological defects
(like cosmic strings) played a role in structure formation. In these
topological defect models the $X$ particles could be superheavy gauge
or Higgs bosons, and/or their superpartners.

Alternatively the $X$ particles could exist as (more or less) free
particles in today's Universe. Such particles might have been created
at the end of inflation, when the energy density of the Universe was
more than one hundred orders of magnitude higher than it is
today. Several production mechanisms have been suggested,
e.g. gravitational production due to the rapidly varying metric (which
could work for $X$ particles as heavy as $10^{16}$ GeV) \cite{md16},
or in (inclusive) inflaton decays \cite{md17}. The $X$ particles would
then serve as ``batteries'', storing energy from this extremely violent
early epoch of the Universe and releasing it in our much balmier times.

In order to explain the longevity of such free $X$ particles one has
to postulate that their couplings to ordinary matter are very strongly
suppressed. Some such suppression is actually expected if $X$ resides
in the ``hidden sector'' thought to be responsible for the spontaneous
breaking of supersymmetry \cite{md18}, which by definition only has
$M_{\rm Planck}$ suppressed couplings to the visible sector, although
it is by no means guaranteed that this suppression is sufficient. The
relevant couplings could also be suppressed by (approximate)
symmetries \cite{md19}, or geometrically in brane world scenarios
\cite{md20}.\footnote{It is sometimes claimed \cite{md21} that the
required long lifetime of $X$ needs severe finetuning. This is not
really true, since $\tau_X$ could be as large as $10^{20}$ yrs
\cite{md22} if $X$ particles constitute the bulk of Dark Matter. Model
builders therefore have some ten orders of magnitude in $\tau_X$ to
aim for. Of course, the observed UHECR flux determines the ratio
$\Omega_X / \tau_X$ to within a factor of a few, but this does not
require any more finetuning of model parameters than in any other
explanation of the UHECR events.}

\begin{figure}[htbp]
\epsfxsize=22cm
\vspace*{-4cm} \hspace*{-4cm} \epsfbox{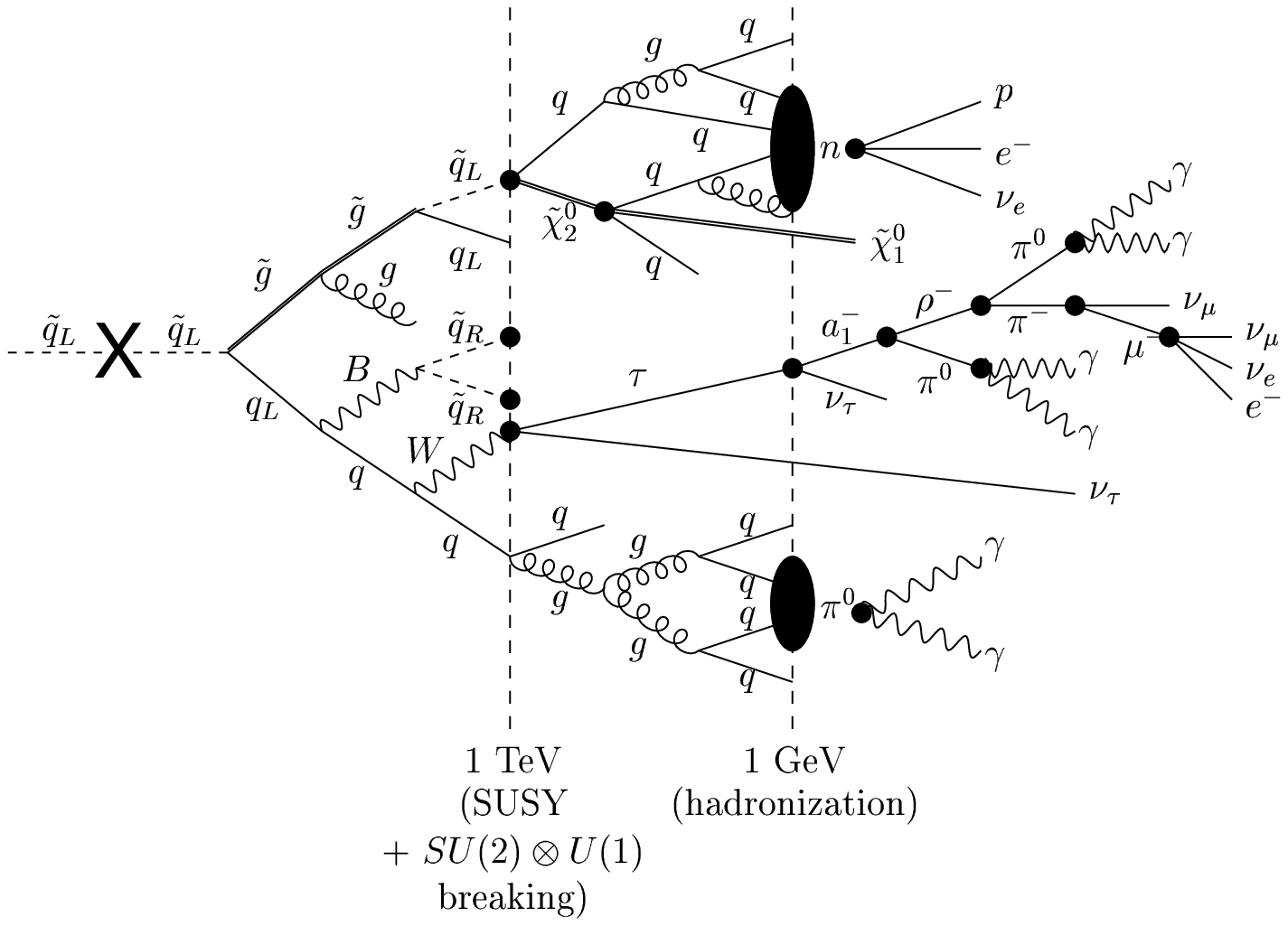} \vspace*{-16cm}
\caption{\small Schematic MSSM cascade for an initial squark with a
virtuality $Q \simeq M_X$. The full circles indicate decays of massive
particles, in distinction to fragmentation vertices. The two vertical
dashed lines separate different epochs of the evolution of the
cascade: at virtuality $Q > M_{\rm SUSY}$, all MSSM particles can be
produced in fragmentation processes. Particles with mass of order
$M_{\rm SUSY}$ decay at the first vertical line. For $M_{\rm SUSY} > Q
> Q_{\rm had}$ light QCD degrees of freedom still contribute to the
perturbative evolution of the cascade. At the second vertical line,
all partons hadronize, and unstable hadrons and leptons decay. See the
text for further details.}
\end{figure}

The spectrum of UHECR primaries (and other stable particles) at source
is determined by the physics of $X$ decays \cite{md23}. This is
quite nontrivial, as indicated in Fig.~1, which is taken from
ref.\cite{md24}. The primary few--body $X$ decay initiates a
generalized parton shower. This is similar to, e.g., the hadronic
decay of $Z$ bosons, which have been studied in great detail at LEP
and SLC. From a theorist's perspective, a $Z$ boson decays into a $q
\bar q$ pair, whereas experimentally one observes twenty or so hadrons
in the final state. The transition from the two (anti)quarks to the
${\cal O}(20)$ hadrons entails both perturbative and non--perturbative
physics. The perturbative part is called a parton shower. It can be
envisioned by assuming that the original $q \bar q$ pair is not
produced on--shell, but with initial time--like virtualities of order
$m_Z$. The showering occurs when these partons move closer to the mass
shell by emitting additional partons, mostly gluons, with smaller
time--like virtualities.
\setcounter{footnote}{0}

Similarly, the particles produced in primary $X$ decay should be
considered to have initial time--like virtualities ${\cal O}(m_X)$. At
such high energy scales parton showers are expected to develop quite
differently from parton showers at scale $m_Z$. To begin with, the
existence of a (real, four--dimensional) energy scale $m_X \gg m_Z$
strongly indicates the existence of superparticles, which can
stabilize this very large ratio of scales against radiative
corrections \cite{md25}. Since the superparticle mass scale $m_{\rm
SUSY} \lsim 1$ TeV $\ll m_X$, superparticles can be produced (i.e.,
will be ``active'') in the initial part of the shower development,
even if $X$ decays primarily into SM particles. Moreover, we know that
at energy scales near $m_X$ all three gauge interactions are of
approximately equal strength. In refs.\cite{md24,md26} we therefore
included all gauge interactions as well as third generation Yukawa
interactions in the description of the parton shower.

When the shower virtuality scale reaches $m_{\rm SUSY}$,
superparticles decouple from the shower and decay; the electroweak
gauge and Higgs bosons, as well as top quarks, decouple at essentially
the same scale. A careful treatment of these decays is mandatory if
one wants to account for the entire energy released in $X$ decays. As
well known, sparticle decays can be somewhat complicated even at the
parton level, involving lengthy decay cascades. We treated these with
the help of ISASUSY \cite{md27}. 

At shower scales between $m_{\rm SUSY}$ and $Q_{\rm had} \sim 1$ GeV
only standard QCD is important for the evolution of the parton shower,
as at LEP. At scale $Q_{\rm had}$ the non--perturbative parton
$\rightarrow$ hadron transition occurs. Most hadrons, as well as heavy
$\mu$ and $\tau$ leptons, eventually decay as well, as indicated in
the figure. At the end one is thus left with seven species of
particles (plus their antiparticles): protons, electrons, photons,
three kinds of neutrinos, and LSPs. 

Out of the multitude of particles produced in a far--away $X$ decay at
most one will be observed on Earth. We therefore ``only'' have to
compute the one--particle inclusive $X$ decay spectra into the seven
stable decay products. In QCD the notion of fragmentation functions
(FFs) has been developed precisely in order to describe
single--particle inclusive spectra \cite{md28}. These FFs depend on
two variables: the energy scale of the shower, here $m_X$, and the
scaled energy $x_P = 2 E_P/m_X$ of the stable decay product $P$ in the
$X$ rest frame. The $x$ dependence of hadronic FFs at some reference
scale has to be taken from data, since non--perturbative effects are
important here; we used the fits of ref.\cite{md29}.\footnote{Some of
these ``input'' FFs are still not well determined, e.g. those starting
from a $b$ or $c$ quark.} The scale dependence of the FFs is described
by the appropriate generalization of the well--known DGLAP evolution
equations \cite{md30}. The decays of superparticles and other massive
partons can also be treated in this framework, as long as their energy
is much larger than their mass, in which case a collinear treatment is
adequate. Finally, at very small $x$ color coherence effects should be
included \cite{md24}.

The main results of this analysis can be summarized as follows
\cite{md26,md24}: \begin{itemize}

\item For small $x \lsim 0.01$, the $\nu_e, \, \nu_\mu, e, \gamma$ and
$p$ fluxes\footnote{All fluxes are summed over particle and
antiparticle, if they are distinct.} have essentially fixed ratios,
independent of the primary $X$ decay mode(s) and of the details of the
SUSY spectrum. The $\nu_\mu$ flux is largest, exceeding the $p$ flux
by a factor $\sim 3$, while the photon and $e \simeq \nu_e$ fluxes
exceed the proton flux by a factor $\sim 2.5$ and $\sim 2$,
respectively. All these fluxes can be described by a simple power law
in this region, $d \Phi / d E \propto E^{-1.4}$. These features
are due to QCD evolution effects. Note that most $\nu_{e,\mu}, \, e$
and $\gamma$ at $x \lsim 0.01$ originate from pion decays, and are
thus subject to the same QCD effects.\footnote{These effects also
imply that the results are largely independent of the necessary
extrapolation of the ``input'' FFs towards small $x$, as long as this
extrapolation conserves energy and has a milder small$-x$ behavior
than $x^{-1.4}$.} On the other hand, the $\nu_\tau$ and lightest
superparticle (LSP) fluxes, which decouple much earlier in the shower,
are model--dependent even at small $x$. Generally they increase more
slowly with decreasing $x$, and thus become subdominant for $x \lsim
0.01$.

\item For $x \gsim 0.1$ the fluxes at source depend very strongly on
the primary $X$ decay mode(s), and less strongly (typically at the
factor of two level) on details of the SUSY spectrum, the relative
ordering of and mass splitting between states being more important
than the overall sparticle mass scale. For all $X$ decays into MSSM
particles the photon flux at source is larger than the proton flux. In
case of hadronic $X$ decays the ratio $\Phi_\gamma / \Phi_p$ reaches a
minimum (slightly above 1) at $x \simeq 0.3$. At larger $x$ it
increases due to the emission of hard photons early in the shower,
illustrating the importance of including electroweak interactions in
the description of the shower. At smaller $x$ this ratio increases
because the ratio of FFs into pions and into protons increases with
decreasing $x$ \cite{md29}. For primary $X$ decays into only weakly
interacting particles the photon flux at $x \gsim 0.1$ exceeds the
proton flux by at least one order of magnitude, since in this case
hadrons can only be produced after two electroweak branching
processes, while photons can be emitted already in the first
branching. The $\nu_\mu$ and $\nu_e$ fluxes also exceed the proton
flux at large $x$; these two neutrino fluxes approach each other at $x
\gsim 0.5$, where most neutrinos come directly from electroweak
branching and decay processes, rather than from pion decays.

\item The LSP flux is important even if the primary $X$ decay only
involves SM particles. For example, Bino--like LSPs carry 2--3\% of
the total energy if $X$ decays into leptons, 5--6\% if $X$ decays into
quarks, 20--30\% if $X$ decays into squarks, and 30--50\% if $X$
decays into sleptons. The percentage is higher for $X$ decays into
$SU(2)$ singlet sfermions, since they have shorter supersymmetric
decay chains, and also shower less (due to the absence of $SU(2)$
couplings).

\end{itemize}

These fluxes are modified by propagation effects. Even if most
relevant $X$ decays occur in the halo of our own galaxy, which is
expected \cite{md31} if $X$ particles can move freely under the
influence of gravity, electrons will loose most of their energy in
synchrotron radiation on galactic $\vec{B}-$fields. Moreover, the
propagation distance is so large that oscillations essentially wash
out all information about the original neutrino flavor dependence,
i.e. approximately equal fluxes of $\nu_e, \, \nu_\mu$ and $\nu_\tau$
will reach Earth independent of the ratios of these fluxes at source.

How can one test this class of models? Given that $m_X$ and the source
density $\propto \Omega_X/\tau_X$ are considered to be free
parameters, it is not surprising that one can fit \cite{md23,md32} the
observed UHECR flux. These models are even compatible with a GZK
spectral break, as indicated by the current HiRes (but not AGASA)
data, if $X$ particles are distributed more or less uniformly
throughout the Universe, as would e.g. be expected if they are
confined to cosmic strings. As mentioned earlier, freely moving $X$
particles are expected \cite{md31} to have a large overdensity ($\sim
10^4$ to $10^5$) in our galaxy, in which case most relevant $X$ decays
would occur at distance well below one GZK interaction length. Even in
this case the HiRes spectrum can be described by $X$ decays, simply by
choosing $m_X$ such that the spectrum cuts off just above $10^{11}$
GeV.

The composition of UHECR primaries is more problematic, however,
Observations indicate that most primaries are protons or heavier
nuclei, not photons. This follows from an analysis \cite{md33} of the
longitudinal shower profile of the most energetic event ever observed;
from the large number of muons in the Haverah Park \cite{md34} and
AGASA \cite{md35} data; and from the absence of a South--North
asymmetry in the AGASA data \cite{md35}, which would be expected for
photon primaries, since photons with $E_\gamma > 10^{10}$ GeV split
into $e^+e^-$ pairs already high in the Earth's magnetosphere. In
contrast, as mentioned earlier, $X$ decay models predict the photon
flux at source to be higher than the proton flux. If most sources are
at cosmological distances, propagation effects may well suppress
$\Phi_\gamma$ at post--GZK energies even more than $\Phi_p$. However,
if most sources reside in the halo of our galaxy, propagation effects
on $\Phi_\gamma$ are expected to be significant only if the galactic
radio background has been underestimated by at least an order of
magnitude. I do not know how (im)plausible this might be.

Fitting the observed UHECR flux to $\Phi_p$ alone increases the
predicted flux of neutrinos and LSPs on Earth by about a factor of
three if $m_X \gg 10^{12}$ GeV; if $m_X \sim 10^{12}$ GeV, the
enhancement amounts to a factor $\sim 2$ for hadronic primary $X$
decays, and to roughly one order of magnitude for leptonic
decays.\footnote{Leptonically decaying $X$ particles with $m_X \lsim
10^{13}$ GeV are thus particularly difficult to reconcile with the
experimental evidence against photons as main UHECR primaries.} Even
without this enhancement the neutrino flux might be detectable by
future experiments \cite{md36a}. One can look for long muon tracks as
well as high--energy particle showers due to NC and $\nu_{e,\tau}$ CC
events in neutrino telescopes like IceCube \cite{md36}. Another
technique, used by the RICE experiment at the South Pole \cite{md37},
is to look for coherent Cherenkov emission of radio waves, which is
mostly sensitive to electromagnetic showers, i.e. to $\nu_e$ CC
events. Finally, future air shower detectors like the Pierre Auger
array \cite{md38} can look for ``horizontal'' air showers, which
originate too deep in the atmosphere to be due to photons or protons,
and for $\tau-$leptons coming out of the Earth at shallow angles and
decaying in the atmosphere. In all cases one has to require $E > 100$
TeV in order to suppress the atmospheric neutrino background.

Some estimated \cite{md32} event rates per year are collected in Table
1, for two extreme values of $m_X$ and two extreme $X$
distributions. ``Galactic'' here means that essentially all relevant
sources are closer than one GZK interaction length. In contrast, if
$X$ particles are distributed homogeneously throughout the Universe,
$X$ decays occuring at distances well beyond one GZK interaction
length contribute to $\Phi_\nu$, but not to $\Phi_p$ (at post--GZK
energies). A homogeneous distribution therefore leads to $\sim 10$
times higher neutrino event rates than a galactic distribution
does. Similarly, $m_X \sim 10^{12}$ GeV leads to an about ten times
higher neutrino event rate than $m_X \sim 10^{16}$ GeV does. The
reason is that, in particular for a galactic distribution of $X$
particles, models with $m_X \gg 10^{12}$ GeV only allow to fit the
very highest energy end of the UHECR spectrum, leading to
significantly smaller fluxes at source. Notice also that the cross
sections for $\nu p$ scattering grow significantly less rapidly than
linearly with energy once $E_\nu \gg 1$ TeV \cite{md40}. Finally, the
ranges in the table indicate the spread between different $X$ decay
models. For $m_X \sim 10^{16}$ GeV all relevant particles are at $x
\ll 1$, in which case this model dependence disappears, as mentioned
above. 

\begin{table}[h]
\caption{\small Number of neutrino events with $E_\nu > 100$ TeV per year
expected in different detectors. See the text for further details.} 
\vspace*{3mm}
\begin{center}
\begin{tabular}{|c|c||c|c|c|}
\hline
$m_X$ [GeV] & $X$ distribution & IceCube & RICE & Auger \\
\hline
$2 \cdot 10^{12}$ & Galactic & 10--30 & 1--4 & 1--3 \\
$2 \cdot 10^{12}$ & Homogeneous & 80--300 & 10--35 & 10--25 \\
\hline
$2 \cdot 10^{16}$ & Galactic & 1 & 0.4 & 0.3 \\
$2 \cdot 10^{16}$ & Homogeneous & 10--15 & 6 & 5 \\
\hline
\end{tabular}
\end{center}
\end{table}
\vskip1pc

Unfortunately the prediction of a detectable neutrino flux at energies
beyond that reached by atmospheric neutrinos is hardly unique for this
explanation of the UHECR events. The very GZK effect itself gives rise
to extremely energetic neutrinos through the decay of charged
pions. Note also that the numbers in table 1 differ by more than two
orders of magnitude, i.e. a large range of neutrino fluxes is broadly
compatible with $X$ decay models. These models generically predict
that the neutrino flux extends to somewhat higher energies than the
proton flux does \cite{md24}. However, testing this prediction would
require a fairly accurate measurement of the neutrino flux at energies
well beyond $10^{11}$ GeV, where the expected event rate in the
experiments covered in table 1 is very low.

An unambiguous test of this kind of model could be achieved by
detecting the flux of very energetic LSPs. In ``bottom--up''
scenarios, where the observed UHECR events are explained by the
acceleration of protons (or heavier nuclei), very energetic
superparticles could only be produced when the accelerated protons (or
nuclei) collide with ambient matter. However, the inclusive cross
section for pion production is at least seven orders of magnitude
larger than that for the production of superparticles. In this kind of
model a detectable LSP flux would therefore be accompanied by a huge
flux of extremely energetic neutrinos, which would be (comparatively)
easily detectable. In contrast, we saw earlier that LSPs are expected
to carry a significant fraction of the energy released in $X$ decays. 

Detecting this UHE LSP flux is difficult, but may not be entirely
hopeless \cite{md41}. The cross section for LSP--nucleon scattering is
estimated \cite{md42} to be 10 to 100 times smaller than the
neutrino--nucleus cross section, if the LSP is Bino--like, which is
the case in most SUSY models. These LSPs will therefore be absorbed
less strongly in the Earth than neutrinos are. For $E_\nu = 10^9$ GeV
only one in a thousand neutrinos impinging on Earth at an angle at
least 5$^\circ$ below the horizon can transverse it without
interacting. $\nu_\tau$s can be regenerated via $\tau \rightarrow
\nu_\tau$ decays even if they undergo CC scattering, but the emerging
$\nu_\tau$ will only carry 20\% or so of the energy of the original
one. If the UHE neutrino ``background'' comes from $X$ decays, one can
therefore show that very few events with $E > 10^9$ GeV can emerge at
an angle $\theta > 5^\circ$ below the horizon, and essentially none at
$\theta > 10^\circ$. In contrast, if $\sigma_{\tilde \chi p} =
\sigma_{\nu p}/10$, at $E_{\tilde \chi} = 10^9$ GeV the LSP flux will
be depleted significantly only at $\theta > 60^\circ$; the depletion
becomes altogether negligible for $\sigma_{\tilde \chi p} =
\sigma_{\nu p}/100$, up to energies well beyond $10^{10}$ GeV.

Due to the LSPs' small flux and cross section, one will need huge
targets. These might be provided by space--borne fluorescence
detectors like EUSO \cite{md43} and OWL \cite{md44}. Table 2 shows
estimated \cite{md41} rates of LSP events with $\theta > 5^\circ$ and
$E_{\tilde \chi} > 10^9$ GeV, assuming $\sigma_{\tilde \chi p} =
\sigma_{\nu p}/10$ and a target area of 150,000 km$^2$. Since LSPs can
interact either in the atmosphere or in $\sim 10$ m water--equivalent
(w.e.) below it, this corresponds to an approximate target size of
2,000 km$^3$ w.e. . Even in that case the expected event rates are not
very large -- and we haven't included any efficiencies yet. Optical
observations of air showers are typically only possible on clear,
moonless nights, i.e. some 10\% of the time. Once this is included,
only the most optimistic scenario (a homogeneous distribution of
$10^{12}$ GeV particles, and an LSP--nucleon scattering cross section
near the upper end of the expected range) will give a clear signal in
this kind of experiment. However, an enlargement of the target area is
at least conceivable, e.g. by simply launching more satellites, or by
using a higher orbit (with correspondingly tougher requirements on the
optics). 

\begin{table}[h]
\caption{\small Number of LSP events with $E_{\tilde \chi} > 10^9$ GeV
emerging from at least $5^\circ$ below the horizon expected each year in
space--borne fluorescence detectors, for $\sigma_{\tilde \chi p} =
\sigma_{\nu p}/10$; a ten times smaller scattering cross section gives
about 8 to 9 times smaller rates. The notation is as in table 1.}
\vspace*{3mm}
\hskip4pc
\begin{center}
\begin{tabular}{|c||c|c|}
\hline
$m_X$ [GeV] & Galactic dist. & Homogeneous dist. \\
\hline
$2 \cdot 10^{12}$ & 2--30 & 30--400 \\
$2 \cdot 10^{16}$ & 0.2--0.6 & 3--10 \\
\hline
\end{tabular}
\end{center}
\end{table}

\section{Summary and Conclusions}

In this contribution I discussed particle physics explanations of the
post--GZK cosmic ray events. My personal favorite remains the
explanation in terms of superheavy late decaying particles, which can
solve both the energy and propagation problems. The biggest current
challenge of this class of models seems to be how to explain the
paucity of photons as primaries in the observed events. A first
minnowing of models should be achieved by measuring the flux of
neutrinos with $E_\nu > 10^5$ GeV. A decisive test may require
measuring (or excluding) the predicted flux of LSPs with $E_{\tilde
\chi} \gsim 10^9$ GeV, which will be difficult, but may be possible
using space--based experiments. Proving the existence of such
mysterious super--heavy but long--lived particles would revolutionize
our understanding of physics. If these models are still viable after
the next generation of cosmic ray observatories and neutrino
telescopes have presented their data, no effort should be spared to
search for these extremely energetic cosmic LSPs.

\subsection*{Acknowledgements}
I thank Cyrille Barbot, Francis Halzen and Dan Hooper for the
collaborations that made this report possible. This work was partially
supported by the SFB 375 of the Deutsche Forschungsgemeinschaft.

\end{document}